\documentclass[12pt,preprint]{aastex}

\usepackage{amsmath}
\slugcomment{To be submitted to ApJ}

\shorttitle{An S$_2$ Cometary Fluorescence Model}
\shortauthors{Reyl\'e and Boice}

\begin{document}

\title{An S$_2$ Fluorescence Model for Interpreting High-Resolution Cometary 
Spectra. I. Model Description and Initial Results.}

\author{C\'eline Reyl\'e}
\affil{Observatoire de Besan\c{c}on, BP 1615, 25010 Besan\c{c}on Cedex, France}
\email{celine@obs-besancon.fr}
\and
\author{D. C. Boice}
\affil{Southwest Research Institute, 6220 Culebra Road, San Antonio, TX 78228}
\email{boice@swri.space.edu}

\begin{abstract}
A new versatile model providing S$_2$ fluorescence spectrum as a function of
time is developed with the aim of interpreting high resolution 
cometary spectra. For the S$_2$ molecule, it is important to take into 
account both chemical and dynamic processes because S$_2$ has a short 
lifetime and is confined in the inner coma where these processes are most 
important. The combination of the fluorescence model with 
a global coma model allows for the comparison with observations of column 
densities taken through an aperture and for the analysis of S$_2$ fluorescence
in different parts of the coma. Moreover, the model includes the rotational 
structure of the molecule. Such a model is needed for interpreting recent
high spectral resolution observations of cometary S$_2$.
A systematic study of the 
vibrational-rotational spectrum of S$_2$ is undertaken, including relevant 
effects, such as non-equilibrium state superposition and 
the number density profile within the coma due to dynamics and chemistry,
to investigate the importance of the above 
effects on the scale length and abundance of S$_2$ in comets. 
\end{abstract}

\keywords{comets: general --- fluorescence --- S$_2$ --- spectroscopy}

\section{Introduction}

Although comets hold important clues to the formation of the solar system, 
there is no complete picture of where comets formed \citep{mumma00,irvine00}.
The mixing ratio of the minor 
constituents of frozen gases in the ice-dust conglomerate of the nucleus
is a very important clue to the original composition of the frozen gases in
the solar nebula, but it is not well understood. A detailed and comprehensive
description of production of minor species, such as S$_2$, and their sources,
and the inner coma chemistry are prerequisites for unambigous characterization
of the nucleus. Understanding comets, in turn, yields insights into the origin
of the solar system.

Several sulfur species have been detected in comets, including S$_2$ which 
has a short lifetime 
and is concentrated in the innermost coma. The proximity of Comet 
IRAS-Araki-Alcock 1983d \citep{ahearn83} and Comet Hyakutake C/1996 B2 
\citep{weaver96,laffont98} to the Earth favored the detection of emission 
lines of such a short-lived species. Recently, S$_2$ has been detected in 
Comet Lee C/1999 H1 \citep{feldman99}.  

Up to now pure vibrational models were used to analyse these data. 
\citet{kim90}, using a multicycle fluorescence model, derived an abundance of 
2.5$\times10^{-4}$ relative to water in International Ultraviolet
Explorer (IUE) spectra of Comet IRAS-Araki-Alcock. The model 
developed by \citet{laffont98} shows that there is a fast evolution of S$_2$ 
fluorescence spectra with the age of the molecule as it undergoes the
solar radiation and consequently with the distance to 
the nucleus. On the basis of this model, it was possible to retrieve an upper 
limit to the abundance of 10$^{-4}$ relative to water from IUE observations of 
Comet Hyakutake which is consistent with the abundance derived from HST 
observations a few days later \citep{weaver96,ahearn99}. 
The abundance of S$_2$ in Comet Lee is 3-5 $\times10^{-5}$ relative to water 
\citep{feldman99}.

Having applied a time-dependent model of the vibrational population of S$_2$ to
the fullest extent, it 
appeared that a more extensive model including the rovibrational structure of 
S$_2$ was needed to provide additional clues about the physical and 
chemical nature of the inner coma of comets, especially since high spectral
and spatial resolution cometary spectra containing S$_2$ exist now. 
In this paper, we decribe such a 
model suitable for interpreting high resolution spectra of S$_2$ and 
we combine the resulting spectra with a global coma model based on gas 
dynamics and chemistry. This is important for cometary applications and make it
possible to follow the time evolution of S$_2$ fluorescence in the inner coma.

Section \ref{model} describes the model and presents our method of calculation. 
Synthetic S$_2$ spectra at various age of the molecule are presented 
in section \ref{results}.
In section 
\ref{comparison}, we compare this model with the pure vibrational model and 
point out important consequences of using a pure vibational model for
analysing spectra of even moderate resolution. We apply our model to such 
observations, the IUE spectra of Comat IRAS-Araki-Alcock in which S$_2$ was 
first detected and discuss the effects on the abundance and lifetime 
determination. In section \ref{application}, we explore the possible 
applications of this model for cometary observations, both with high spatial
and spectral resolution. Finally, we summarize our conclusions and give 
directions for futur model enhancements in section \ref{conclusion}.

\section{Model Description}

\subsection{Time-Dependent Fluorescence Model}
\label{model}

The $B^3\Sigma^-_u$-$X^3\Sigma^-_g$ system of S$_2$ has been
analysed in detail in the laboratory (see, e.g. 
\cite{ikenoue60,heaven84,matsumi84,smith81}) but, up to now, there is no complete 
model of S$_2$ rovibrational fluorescence. Previous models only took into acount 
the vibrational levels of the molecule. \citet{kim90} used a multi-cycle fluorescence 
approach whereas our previous model \citep{boice97,laffont98} gave
the fluorescence spectrum at any time after the release of the molecule from 
the nucleus. This time-dependent method is quantitatively different from the 
multi-cycle fluorescence approach. At no time do we assume that a given level 
is in steady state. This allows us to calculate the spectrum of the molecule
at any moment. The period of time elapsed since the molecule release can be
very short, of the order of microseconds, or very long, up to the point where 
steady state is reached. An immediate difference is noted between our results
and those of \cite{kim90}. They argue that the ``first-cycle'' spectrum is 
achieved only after about 100~s but we find that this spectrum is obtained after
a few seconds.

We now have enhanced our time-dependent model by considering the rotational 
levels. We calculate synthetic fluorescence spectra of the 
$B^3\Sigma^-_u$-$X^3\Sigma^-_g$ system of S$_2$ by solving the 
evolution equation of the population $n_i$ of the 
rovibrational levels $i$ as a function of time:
\begin{equation}
\label{population}
\dot{n}_i\ = - n_i \sum_j P_{ij} + \sum_j n_j P_{ji} + Q_i \hspace{.3cm}i=1,...,N
\end{equation}
where $P_{ij}$ is the transition probability between levels $i$ and $j$. 
$P_{ij}$ may 
be either $A_{ij}$ in emission or $B_{ij} \rho_{ij}$ in absorption. Stimulated
emission was considered and found to be negligible. The 
Einstein coefficients $A_{ij}$ and $B_{ij}$ were 
calculated using the Franck-Condon factors from \citet{anderson79}, the 
absolute lifetimes from \citet{quick81}, and the solar irradiance $\rho_{ij}$ 
from \citet{ahearn83b}. The H\"onl-London factors for $\Sigma$-$\Sigma$ 
transitions are derived from \citet{herzberg89} formulae. There is no Q-branch 
for such transitions. 
In addition to the main branches P and R, satellite branches can also 
appear with appreciable strength \citep{tatum71,meyer73}. At the moment, we 
did not consider satellite branches in our calculations, but we are aware that 
this may cause discrepencies with observations at high spectral resolution.

The model was constructed to include a general source (sink) term, 
$Q_i$, for taking into account processes such as collisional excitation 
(deexcitation).  An estimate of the role of collisions shows that they are 
likely to be important very close to the nucleus. At a  distance of 3 km 
above the nucleus surface, the number density of water (the dominant gas) is 
roughly $10^{13}$ cm$^{-3}$.
This results in a collision frequency of 350 s$^{-1}$, assuming a collisional 
cross section of $10^{-15}$ cm$^{-2}$. Collisional processes may dominate the 
excitation in this region. At a distance of 400 km which is the 
approximate range of the observed S$_2$ emissions, the collision frequency is 
reduced to 8 $\times 10^{-2}$ s$^{-1}$, but still comparable to the excitation 
rate by the solar radiation. The 
observed S$_2$ intensity integrated along the line of sight 
will contain a contribution from the region near to the nucleus where 
collisions may be very important and a contribution from further out where 
collisions may also influence the emission process, but in a more moderate 
way. The relative weight of 
these two types of contributions will require a detailed investigation beyond 
the scope of the present study. It should be kept in mind that collision rates of 
S$_2$ with other molecules are unknown at the low temperatures typical of the 
coma. Additional uncertainty in the model parameters would be introduced in 
order to include these collisional effects. Since the primary aim of this 
study is to calculate high-resolution, ro-vibrational synthetic spectra 
excited by solar fluorescence, collisional processes are outside the scope of 
these calculations. In a subsequent study, collisional processes will be 
investigated within the framework of the model outlined above.

We calculated the lines positions 
using the molecular constants for the $X^3\Sigma^-_g$ and $B^3\Sigma^-_u$ 
electronic states of S$_2$ listed by \citet{herzberg89}. Molecular 
constants obtained from observations \citep{ikenoue60} give no significant
difference.
We took into account the Swings effect due to the heliocentric velocity 
of the comet. The initial distribution is obtained with a Boltzmann temperature
of 200 K, which is a typical temperature of the expanding gas at the surface 
of the nucleus, predicted by coma models \citep{boice96}.

The model S$_2$ molecule contains 34 vibrational levels in the electronic ground 
state $X$ and 10 in the excited state $B$. The molecule predissociates above 
this level \citep{ricks69}. We considered hundred rotational levels within 
each vibrational level. 
Since deexcitations due to collisions have been neglected, no 
constraints are imposed on the rotational temperature. We decided to use a 
large number of rotational levels even though the higher ones are not 
effectively populated.
The nuclear spin of S$_2$ being equal to zero, the anti-symmetric levels
are not populated. Given the selection rules, we considered more than 2000 
rovibrational levels. The problem is to solve the 
large number of rate equations in the 
coupled network of ordinary differential equations, subject to the constraint 
of a conservation equation. Therefore, the problem is overdetermined and one 
of the equations is redundant. In practice, we solve the complete set of 
equations and use the conservation equation $\sum_i n_i = 1 $ to check the 
accuracy of the solution. The system is solved using the Gear method 
\citep{gear71} for stiff 
differential equations where time constants vary by many orders of magnitude. 
This method uses variable time steps and error control techniques to preserve 
accuracy during the integration. Global linear invariants are found to be 
kept constant approximately to machine accuracy.

\subsection{Coma Chemistry Model}
\label{chimie}
We have developed a global model for the reactive multifluid gas flow in 
the coma, including gas release from the nucleus, the relationship of
chemical abundances in the coma to those in the nucleus ices (depending on
heliocentric distance of the comet), entrainment of dust by escaping coma 
gases, and fragmenation and evaporation of dust (distributed coma sources)
\citep{huebner91,boice95}. The model solves the fluid dynamic equations for
the mass, momentum, and energy for three neutral fluids (atomic and molecular
hydrogen and the heavier bulk fluid.), and the elctrons. In the inner coma, 
the gas expands, cools, accelerates, and undergoes a plethora of photolytic
(with optical depths effects) and gas-phase chemical reactions that transform
the dozen or so parent molecules into hundreds of daughter species. Beyond the
collision region, the multifluid gas flow for fast atomic hydrogen, fast
molecular hydrogen, electrons, and the bulk of the coma gas undergoes a
transition from fluid dynamic flow to free molecular flow. 

The model 
produces cometocentric abundances of the coma gas species; velocities of the
bulk gas, light atomic and molecular hydrogen with escape, and electrons; gas
and electrons temperatures; column densities to facilitate comparison with
observations, coma energy budget quantities; attenuation of the solar 
irradiance; and other quantities that can be related readily to observations.
Model results are integrated in the line of sight for direct comparison with 
observations.

Several mechanisms have been proposed to explain the presence of S$_2$ in the
coma of comets \citep{vanysek93,saxena95,ahearn00}. For our initial 
model we assume the simplest
possible case: direct release of S$_2$ from the nucleus with a production rate
of 10$^{-4}$ that of water, in order to simulate observations of Comet 
Hyakutake \citep{laffont98,ahearn99}. The S$_2$ molecule
photodissociates at one of three assumed rates: 100s \citep{meier99}, 200s 
\citep{dealmeida86,budzien92}, and 450s \citep{ahearn83}, 
as this rate is not well known. 
No dust was included for this
work.
With such simple chemistry, 
we have used our coma model to incorporate dynamical and optical depth
effects accurately since S$_2$ is confined to the innermost region of the
coma where these effects are most important. 

\section{Results}

\subsection{Synthetic Fluorescence Spectra}
\label{results}

\begin{figure}
\plotone{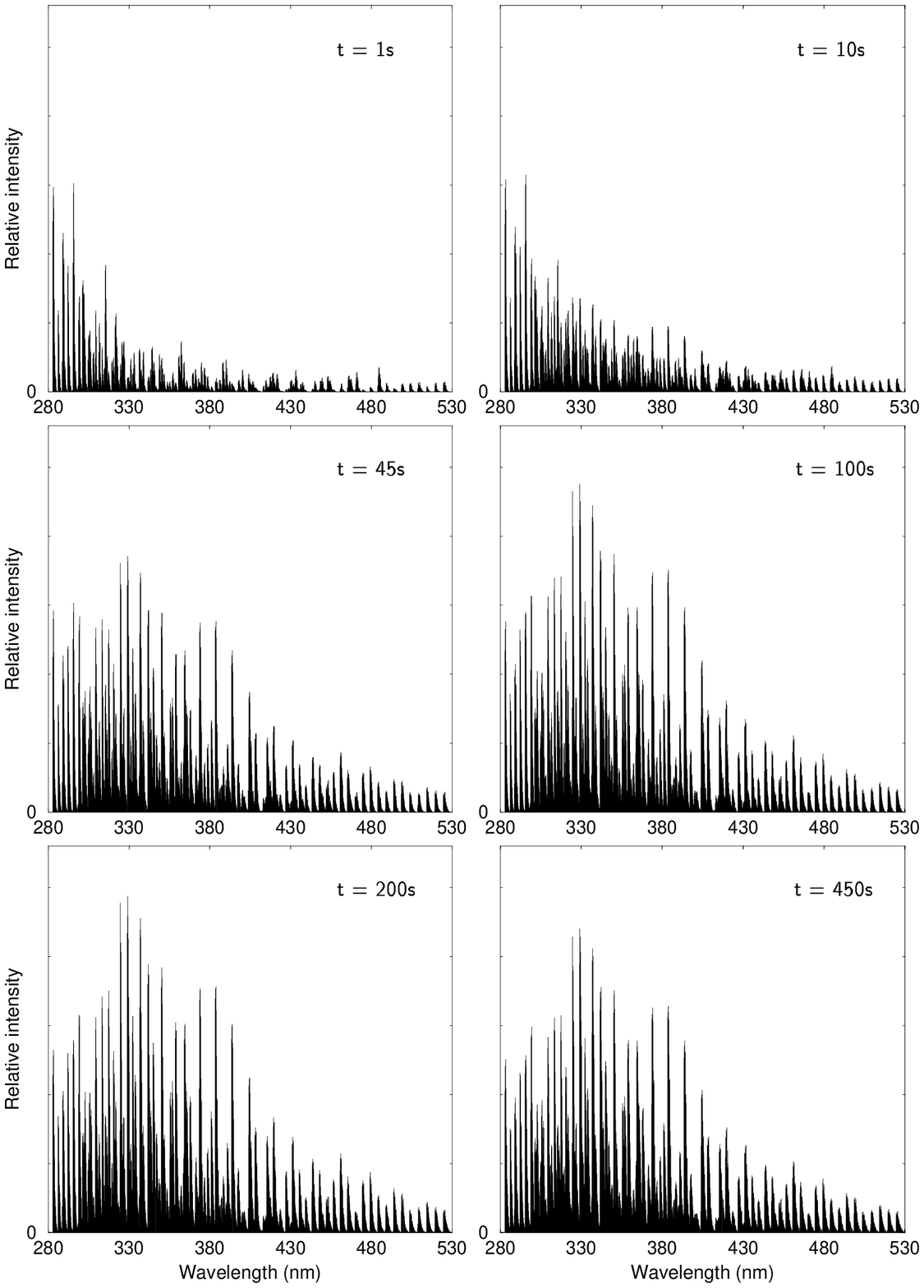}
\figcaption[f1.eps]{Synthetic fluorescence spectra of S$_2$ for 
molecule exposed from $t$ = 1 to 450~seconds to the solar radiation for a 
1~AU heliocentric distance. \label{f1}}
\end{figure}

\begin{figure}
\plotone{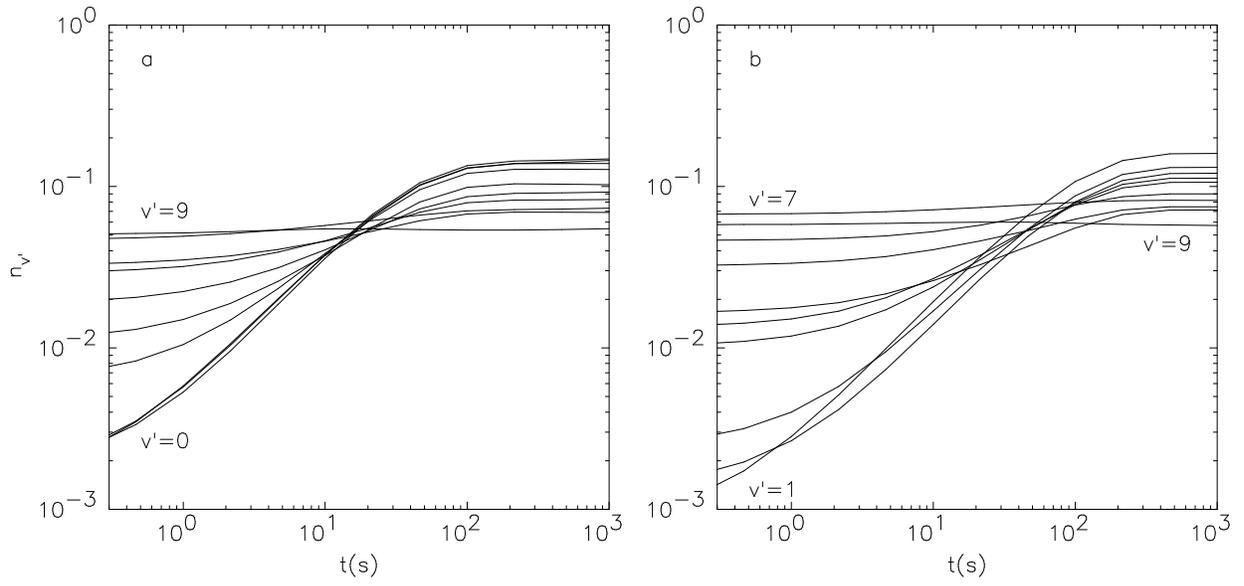}
\figcaption[f2.eps]{Evolution with age of the molecule of the 
population of the 
vibrational levels in the electronic B-state. (a) Rovibrational model. 
(b) Pure vibrational model. \label{f2}}
\end{figure}

Individual synthetic fluorescence spectra of S$_2$ are presented in 
figure~\ref{f1} as a function of age of the molecule, obtained 
considering an irradiation equivalent
to the solar irradiation at a heliocentric distance of 1~AU. At short 
times, corresponding to small cometocentric distances, the brightest lines are 
in the range 280 - 300~nm (e.g the 9-0 band at 
282.9~nm and the 7-1 band at 296.1~nm) and the intensity of
these lines does not change a lot with age $t$.
These lines correspond to transitions from the levels $v^\prime$ = 7 to 9 that
reach a value close to the steady state value within 1s, as shown on 
figure~\ref{f2}a. This figure represents the population of the vibrational
levels in the B-state, that is the sum of the population of the rotational
levels in one vibrational level. When the time increases,
the lower vibrational levels are populated and the lines at long wavelengths
become brighter, such as the 3-3 band at 324.5~nm or the 2-3 band at 328.9~nm. 

After a period of time of roughly 100s, the relative intensities of the 
lines are nearly constant (see figure~\ref{f1}). But inside a vibrational 
level, the population of the higher rotational levels continue to slowly 
increase, approaching steady state with the solar irradiation, 
to the detriment of the lower rotational levels.

\subsection{Comparison with the Pure Vibrational Model}
\label{comparison}

Our pure vibrational, time-dependent model tended to show that the fluorescence 
spectrum varies the most between t~=~0 and 200~seconds, when the intensities 
have reached 90\% of the steady-state values, but very little afterwards 
until about 600s, when steady state is effectively achieved 
(figure~\ref{f2}b). This result is 
different from the one deduced from our rovibrational model which shows that 
the global population of the vibrational levels are constant after about 100s.
The numerous rotational levels allow the electrons to reach higher levels at a 
faster rate due to the overestimate of the Einstein coefficients for
the emission when ignoring the rotational levels. 

Using a pure vibrational 
model to interpret the relative intensities can lead to errors in estimating 
the effective solar irradiation time of S$_2$ in the coma which can affect its derived 
lifetime against destructive processes. Indeed, the S$_2$ lifetime is not a 
well known parameter. The lifetime deduced from observations in Comet 
IRAS-Araki-Alcock \citep{ahearn83,budzien92} and in Comet Hyakutake 
\citep{meier99}, and from laboratory studies \citep{dealmeida86} ranges 
between 100 and 450~seconds. The last 
value was derived by \citet{ahearn83} and is consistent with the fact that the 
steady-state spectrum
gives the best fit for Comet IRAS-Araki-Alcock. However, we have shown above 
that even if considering a lifetime as short as 100s, the molecules 
have enough time to reach a state closely resembling the steady state. 
Lets consider a lifetime of 200s which is in better agreement with the 
laboratory calculations \citep{dealmeida86} and the scale length of 200~km 
derived more recently from the IUE spectrum of Comet IRAS-Araki-Alcock 
\citep{budzien92}. The g-factors being very similar to those of the 
steady-state, the number of molecules in the coma will be the same and the
abundance will be roughly twice higher than when considering a 450~s lifetime.
An overestimate of the lifetime leads to a smaller value of derived abundance.
Care must be taken when comparing abundance values derived by
different groups.

Another difference between the two models is the displacement between the 
band heads considered in the pure vibrational model and the wavelength of 
maximum intensity resulting from the development of the bands over many 
rotational levels. Once the rovibrational lines are convolved with a FWHM 
gaussian of 1.1 nm corresponding to the instrumental resolution of IUE,
the displacement is around 3~\AA, that is similar to the shift
observed in IUE spectra of Comet IRAS-Araki-Alcock \citep{ahearn83}.

\subsection{Application to Cometary S$_2$}
\label{application}

\subsubsection{S$_2$ Fluorescence in a Column of the Coma}
\begin{figure}
\plotone{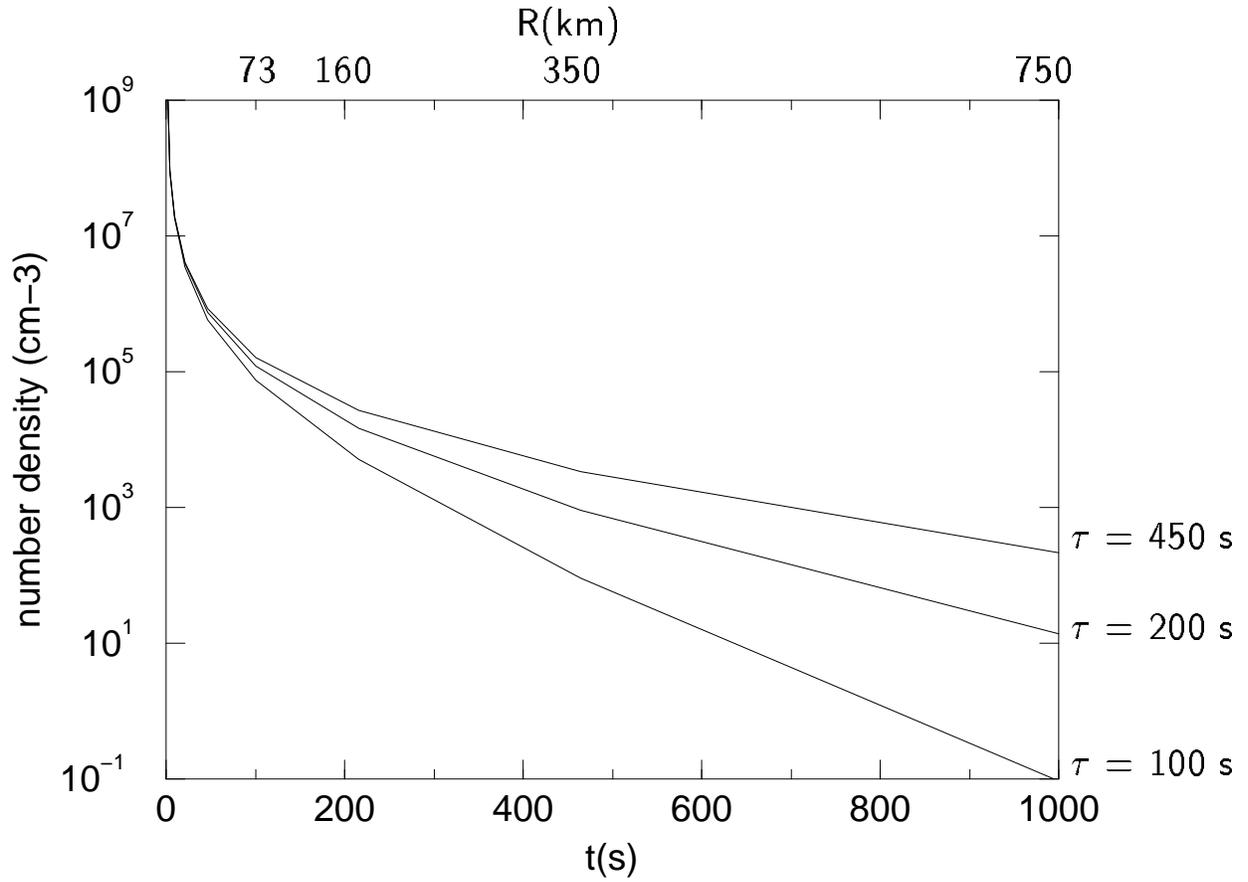}
\figcaption[f3.eps]{Number density of S$_2$ molecules versus the age
of the molecule and the distance to the nucleus given by the model based on gas dynamics
and chemistry for three values of the lifetime: $\tau$ = 100, 200 and 450~s.
\label{f3}}
\end{figure}

An observation of a given region of the coma gives the superposition of 
emissions of all the molecules in a column. These molecules are at different 
cometocentric distances, they have different age $t$ and hence have 
undergone different solar
irradiation times. In order to
model the S$_2$ cometary fluorescence, we combined our fluorescence model with
a global coma model based on dynamics and chemistry as described in section
\ref{chimie} \citep{huebner91,boice95,boice96}. The 
number density $n$ of S$_2$ given by the coma model is shown 
in figure~\ref{f3} as a function of age $t$ and distance to the 
nucleus surface $R$ for different lifetimes: $\tau$ = 100s, 200s and 450s. 

We weighted the individual fluorescence spectra with
the number density given by the coma model as follows:
\begin{equation}
\label{combinaison}
\overline{F}_\lambda(t,\rho) = \int_{-\frac{\pi}{2}}^{\frac{\pi}{2}}
F_\lambda(t) n(t) 
\frac{\rho}{cos^2\theta}d\theta
\end{equation}
where $F(t)$ is the synthetic spectrum obtained at the time $t$ since 
formation of the molecule, 
$n(t)$ is the number density at the distance $R$ from the nucleus surface 
reached by the molecule in $t$, and $\rho$ is the projected distance of the 
column to the nucleus surface. The angle $\theta$ varies from $-\frac{\pi}{2}$
to $\frac{\pi}{2}$ to include the whole column.
We have ignored the Greenstein effect for molecules within the column.

\begin{figure}
\plotone{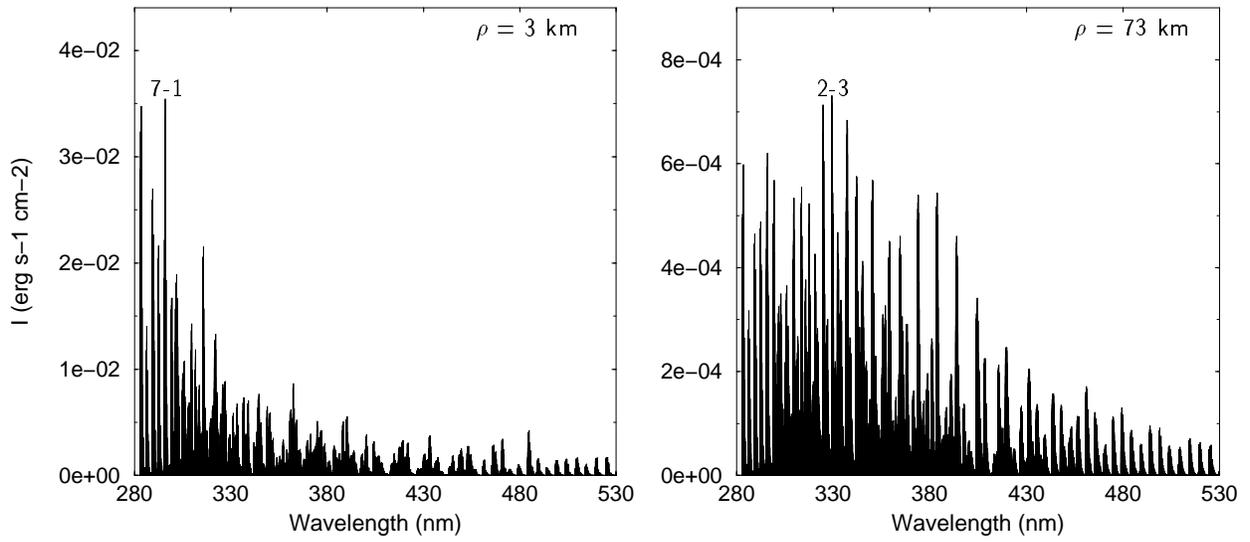}
\figcaption[f4.eps]{Simulation of the fluorescence emission of S$_2$ molecules
within a column very close to the surface nucleus and 73~km away from the
nucleus. The bands of maximum intensity are indicated in both cases. 
\label{f4}}
\end{figure}

The superimposed non-equilibrium spectra that would be
detected on the surface of the nucleus and 73~km away from it are plotted in
figure~\ref{f4}, using a lifetime of 200~s. The number density decreases by more than 10$^3$ from the
nucleus to 73~km away which causes an important decrease in the emission 
intensity. Moreover, we observe a shift in intensity from the $\simeq$ 
280-300~nm region to the $\simeq$ 330-380~nm region. The maximum of emissions 
on the nucleus is the 7-1 band whereas at 73~km away from the nucleus, the 
maximum of emissions has shifted to the 2-3 band. Also, bands at longer 
wavelengths are stronger relative to the brightest bands. 

Up to now, we considered an initial temperature of 200 K, which is a 
correct assumption if S$_2$ is released from the nucleus. Other mechanisms for 
the formation of S$_2$ have been proposed, involving chemical reactions 
\citep{saxena95,ahearn00}, or a grain source \citep{vanysek93}. For these 
alternative assumptions, the excitation of S$_2$ when 
it is formed is different. However, if we consider an initial temperature of 
400 K, the effects due to the initial distribution is not perceptible when 
integrating over the line of sight as soon as one observe a few kilometres 
away from the nucleus.

\begin{figure}
\plotone{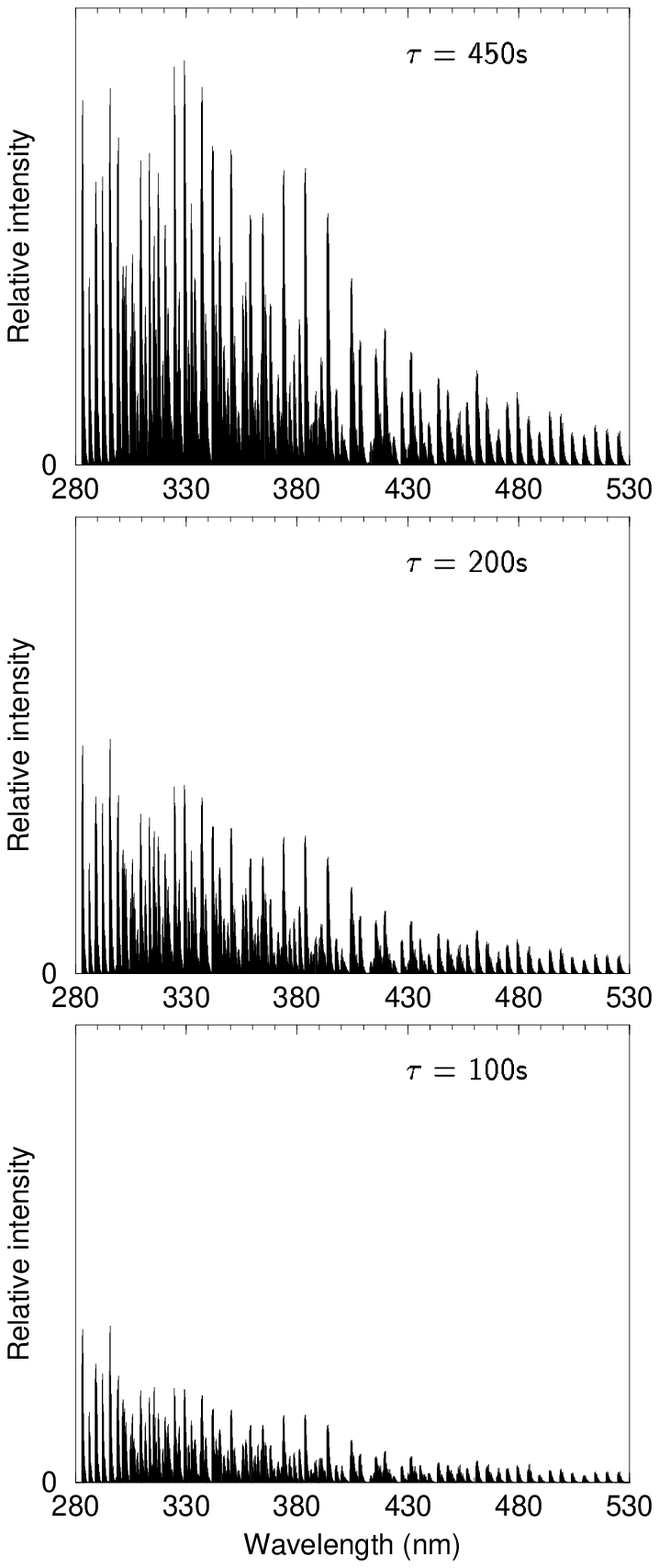}
\figcaption[f5.eps]{Simulation of the fluorescence emission of S$_2$ molecules
within a 68 $\times$ 580~km slit centered on the nucleus considering the
lifetimes $\tau$ = 450, 200 and 100s.
\label{f5}}
\end{figure}

Practically, we do not observe only one column in the coma but many within an 
aperture $A$, for instance a slit or a hole: 
\begin{equation}
{\cal{F}}_\lambda(t,A) = \iint_{\mbox{area}}\overline{F}_\lambda(t,\rho) dA
\end{equation}
It is important to integrate the 
weighted spectra within an aperture for simulating real observations. The
spectrum will be different depending on the width of the slit, or if the
slit is centered or off-centered from the nucleus. For
instance, we can simulate an observation with a slit having an aperture of 
0.87 $\times$ 7.4" or 68 $\times$ 580 km$^2$ centered on the nucleus. This case
actually simulates 
ground-based observations of Comet Hyakutake by \citet{meier99}. 
The simulated spectra are shown in figure~\ref{f5} for the lifetimes 
$\tau$ = 450, 200 and 100s. Considering a short lifetime leads to a smaller 
intensity but the shape of the spectrum is also different depending on $\tau$
(in particular, the relative intensities between the 280-310~nm region and the
rest of the spectrum vary considerably).
This fact is not relevant for spectra such as IUE spectra where the maximum 
wavelength is about 320~nm but may be interesting to consider in the case of
visible spectra such as the ones obtained by \citet{meier99}.
We also considered the effect of having the same slit off-centered with an 
offset of 2''. \citet{ahearn99} reported that with such an offset, S$_2$ was 
no longer detected. With $\tau$ = 450s, the intensities are reduced by a factor
of $\simeq$~3 compared to observations with a slit centered on the nucleus. 
This factor is 5 with $\tau$ = 200s and 12 with $\tau$ = 100s.

\subsubsection{S$_2$ Fluorescence at High Spectral Resolution}
\begin{figure}
\plotone{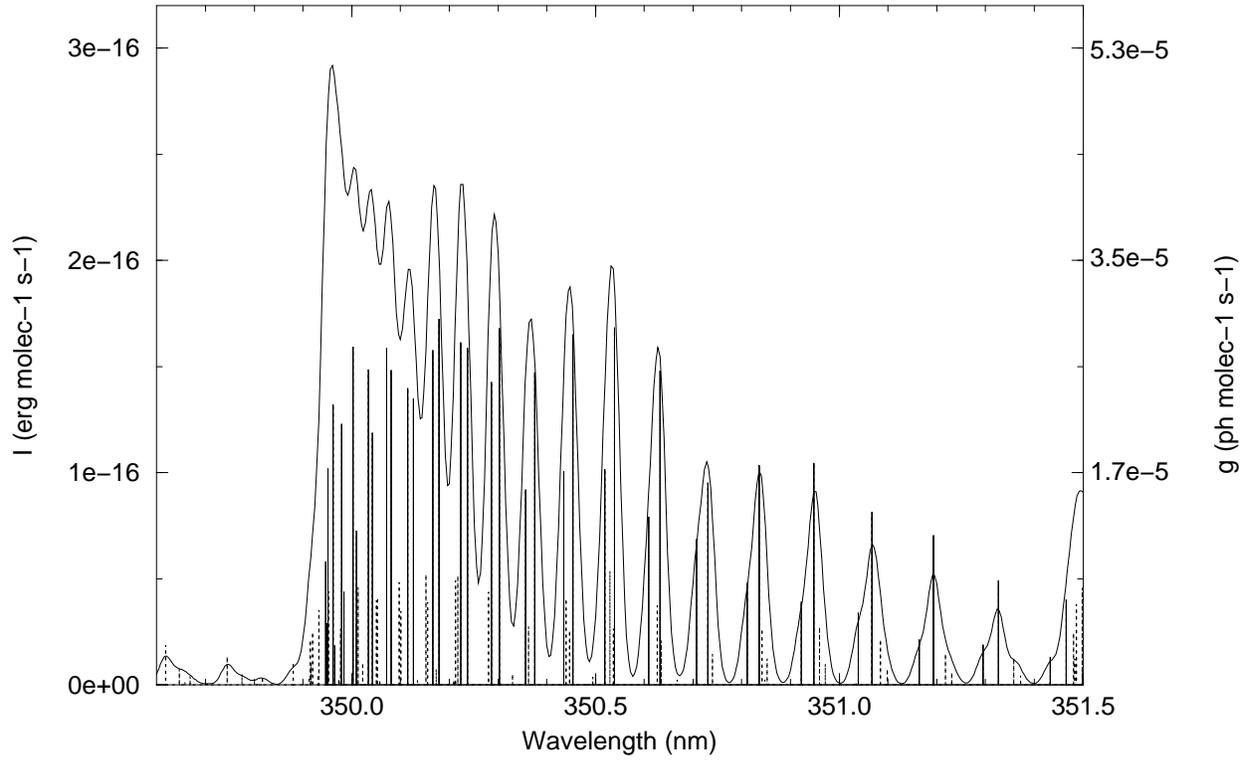}
\figcaption[f6.eps]{Synthetic rotational structure of the 1-5 band (solid 
line) superimposed to the faint 6-8 band (dotted line) for a 200~seconds 
solar irradiation time at 1~AU from the Sun. The lines are convolved with a FWHM 
gaussian of 0.2~\AA~that corresponds to ground-based observations of Comet
Hyakutake \citep{meier99}. The left hand scale is related to the lines 
intensity before convolution. The right hand scale gives the g-factor, or 
fluorescence efficiency, of the lines.
\label{f6}}
\end{figure}

\begin{figure}
\plotone{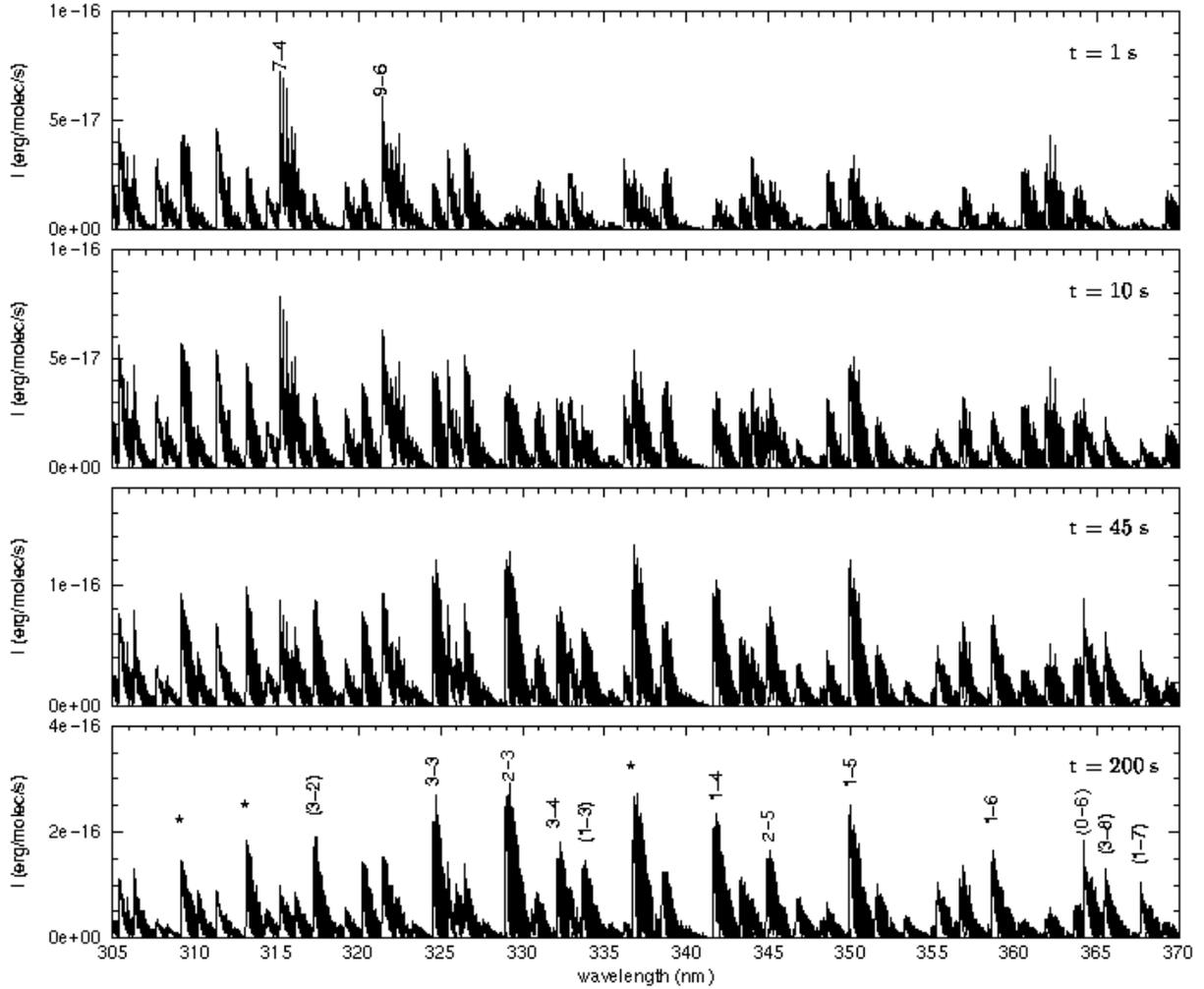}
\figcaption[f7.eps]{Evolution with age of the molecule of the S$_2$ bands at 1~AU
from the Sun. The 
theoretical lines are convolved with a 0.2~\AA~FWHM gaussian. The two
bright bands in the upper graph are characteristic of newly released S$_2$. 
The bands detected in Comet Hyakutake \citep{meier99} are indicated in the 
lower graph. The model shows that these bands are already predominant after 
45~seconds of solar irradiation. The bands in parenthesis may also have been detected 
but were weaker. Bands contaminated by stronger cometray emissions, such as OH 
and NH, are indicated.
\label{f7}}
\end{figure}

Another application of this model is the interpretation of high spectral 
resolution spectra. Figure~\ref{f6} shows the synthetic rotational 
structure of the 1-5 band for a molecule of age $t$=200s. 
This band is 
superimposed with the faint 6-8 band (dotted line). Also shown are the modeled
lines convolved with a FWHM gaussian of 0.2 \AA, which corresponds to 
ground-based observations of Comet Hyakutake \citep{meier99}. Near the band 
head, the convolution shows a steep increase due to the overlapping of many 
lines. Further away, one stucture contains the P and R lines. 
This 1-5 band 
was the brightest one detected by \citet{meier99}. Table~\ref{tab2} lists the
most intense bands predicted by the model after 200s solar 
irradiation time. These bands were effectively detected by \cite{meier99}. The 
model shows that they become dominant very quickly once the molecule 
is released from the nucleus, within 45~seconds (figure~\ref{f7}).
Some of the other brightest bands in table~\ref{tab2} were not detected 
because they are contaminated by stronger emission lines such as bands of 
OH(0-0), OH(1-1), and NH(0-0). Due to no constraint on temperature, the 
rotational temperature given by the modeled spectrum is too high compared to 
the 45K observed rotational temperature \citep{kim00}. 

The upper spectrum in figure~\ref{f7} is obtained for a molecule
aged of 1s.
The two bands, 7-4 and 9-6, are dominant. The intensity of these bands,
particularly the 7-4  band, decrease relative to the other bands when the 
exposure time increases. The dominance of these bands in a spectrum with high 
spatial resolution is therefore a strong indication of freshly released S$_2$ 
molecules.

\section{Conclusion}
\label{conclusion}
We developed a time-dependent fluorescence model of S$_2$ and combined it with 
a global coma model, for interpreting observations in different 
coma regions. As it takes into account the rotational structure of the 
molecule, it can also be used to analyze high spectral resolution observations,
such as those obtained in Comet Hyakutake. We point out that it is also 
important to consider the rotational levels of S$_2$ in the 
fluorescence calculations to interpret even moderate resolution spectra such 
as IUE spectra. We also demonstrated the need of a time-dependent model to
allow better estimates of the lifetime and to determine the
lines characteristic of the freshly released molecules. 

In this work, we concentrated on linking the physics and chemistry of the coma 
and the
cometary fluorescence. This is the first time to our knowledge that a gas
dynamics model with chemistry has been coupled with a time-dependent fluorescence
model to analyse cometary emissions, representing a leap forward in our ability
to gain insight into the physical and chemical processes in the inner coma and
the nature of parent species. This global model with many detailed processes
in the inner coma has great potential to resolve issues regarding the detailed 
rotational lines intensities as functions of distance from the nucleus, 
collision frequency, and production rates and to gain insights into the 
composition into the cometary nucleus. To this end, the future direction of 
this work will be to investigate coma temperature profiles, the effects due 
to optical depth in the inner coma for the emergent S$_2$ emissions, and
also the role of collisions.
Considering collisions would allow self-consistent calculations of the 
rotational temperature for direct comparison to observations. Using the global 
coma model, we will also consider a 
more complete chemistry of S$_2$, not only photolysis for the production and
destruction of S$_2$ in cometary com\ae.

\acknowledgments

We wish to acknowledge support from the NSF Planetary Astronomy Program and the
Observatoire de Besan\c{c}on. We thank Guy Moreels for fruitful comments on the
work and the manuscript.

\clearpage



\clearpage

\begin{table}
\begin{center}
\caption{Most intense bands predicted by the model after a solar
irradiation of 200s. $\lambda_{origin}$ is the wavelength of the band origin. The relative 
intensity are integrated within the band once the lines are convolved with a
0.2~\AA~FWHM gaussian corresponding to observations of Comet Hyakutake from
Kitt Peak Observatory \citep{meier99}. \label{tab2}}
\begin{tabular}{ccc}
\tableline\tableline
Band		&$\lambda_{origin}$(nm)  &relative intensity\\
v$_B$-v$_X$   &&\\
\tableline
1 - 5\tablenotemark{a}  &349.9          &1.00\\
2 - 4\tablenotemark{b}  &336.7          &0.98\\
2 - 3\tablenotemark{a} 	&328.9          &0.97\\
1 - 4\tablenotemark{a}	&341.6          &0.90\\
3 - 3\tablenotemark{a}  &324.5          &0.72\\
1 - 6\tablenotemark{a}	&358.6          &0.69\\
3 - 4\tablenotemark{a}  &332.1          &0.66\\
2 - 5\tablenotemark{a} 	&344.8          &0.64\\
4 - 2\tablenotemark{c}  &313.1          &0.62\\
3 - 2\tablenotemark{a}	&317.2          &0.61\\
0 - 6\tablenotemark{a} 	&364.2          &0.61\\
1 - 3\tablenotemark{a} 	&333.6          &0.57\\
5 - 2\tablenotemark{d}  &309.2          &0.53\\
\tableline
\end{tabular}
\tablenotetext{a}{Bands detected in Comet Hyakutake \citep{meier99}}
\tablenotetext{b}{Overlapping with NH(0-0)}
\tablenotetext{c}{Overlapping with OH(1-1)}
\tablenotetext{d}{Overlapping with OH(0-0)}
\end{center}
\end{table}

\end{document}